\documentclass{article}

\usepackage{PRIMEarxiv}
\usepackage{float}
\usepackage[utf8]{inputenc} 
\usepackage[T1]{fontenc}    
\usepackage{hyperref}       
\usepackage{url}            
\usepackage{booktabs}       
\usepackage{amsfonts}       
\usepackage{nicefrac}       
\usepackage{microtype}      
\usepackage{lipsum}
\usepackage{fancyhdr}       
\usepackage{graphicx}       
\usepackage{authblk}
\graphicspath{{media/}}     

\title{The high-dimensional phase diagram and the large CALPHAD model
}

\author[1]{Zhengdi Liu}
\author[2]{Xulong An}
\author[1]{Wenwen Sun\thanks{Corresponding author at: Department of Materials Science and Engineering,
Southeast University, Nanjing 211189, China
Email Address: swwcsu@live.cn}\textsuperscript{*,}}

\affil[1]{Jiangsu Key Laboratory for Advanced Metallic Materials, School of Materials Science and Engineering, Southeast University, Nanjing 211189, China}
\affil[2]{School of Materials Science and Engineering, Changzhou University, Changzhou 213164, China}

\begin{document}
\maketitle

\begin{abstract}
When alloy systems comprise more than three elements, the visualization of the entire phase space becomes not only daunting but is also accompanied by a data surge. Addressing this complexity, we delve into the FeNiCrMn alloy system and introduce the Large CALPHAD Model (LCM). The LCM acts as a computational conduit, capturing the entire phase space. Subsequently, this enormous data is systematically structured using a high-dimensional phase diagram, aided by hash tables and Depth-first Search (DFS), rendering it both digestible and programmatically accessible. Remarkably, the LCM boasts a 97\% classification accuracy and a mean square error of $4.80\times10^{-5}$ in phase volume prediction. Our methodology successfully delineates 51 unique phase spaces in the FeNiCrMn system, exemplifying its efficacy with the design of all 439 eutectic alloys. This pioneering methodology signifies a monumental shift in alloy design techniques or even multi-variable problems.
\end{abstract}


\section{Introduction}
In the intricate world of alloy design, the CALPHAD (Calculation of Phase Diagrams) method has long served as a foundational pillar, shedding light on the complex phase statuses across varied compositions and temperatures for decades(1).

Historically, alloy design has largely been reliant on binary or ternary phase diagrams(2–4). While these diagrams have been invaluable, they exhibit a distinct limitation: their inability to adequately address Complex Concentrated Alloys (CCAs), constrained by the representation of just two compositional variables per figure(5). This shortcoming necessitated the adaptation of algorithmic approaches for CCAs design(6–8). However, these algorithm-driven approaches, like the genetic algorithm, tend to provide a non-systematic design, often gravitating towards local optima and limited compositions that satisfy the criteria. The essence of their inefficiency boils down to a dual problem: the inherent limitations of computational speed and a point-by-point validation approach that inherently lacks systematic exploration.

To truly revolutionize alloy design, one must address both the speed and the systemic design hurdles. The advent of GPU-accelerated machine learning offers a tantalizing solution for the speed conundrum(9, 10). Yet, existing models, with their narrow task-specific architectures such as discerning between FCC or BCC phase statuses, fall short of the holistic needs of alloy design(11–13). In this study, we introduce the Large CALPHAD Model (LCM), representing a manifestation of Artificial General Intelligence (AGI) within the CALPHAD , specifically tailored for combination of multi-classification and regression task. Trained on an extensive dataset and with a custom neural network, complete with specialized activation and loss functions, this model surpasses the limitations of its predecessors by providing predictions that span the entirety of an alloy's phase space. Not only does LCM function at the caliber of traditional CALPHAD models in alloy design tasks, but its enhanced speed—approximately 50,000 times faster—positions it as a potential benchmark in GPU-accelerated engineering applications.

With the speed obstacle tackled, the focus pivots to establishing a systematic and comprehensive alloy design strategy. Recalling the intuitive method of seeking overlaps in target phase spaces within the confines of binary phase diagrams, we propel this foundational understanding into a new dimension. Our innovative high-dimensional phase diagram is an evolution of this core principle, meticulously adapted for the complexities of CCAs. For the first time, employing a high-dimensional list as a hash table, a unique mapping between composition, temperature, and list indices is established. Augmented by the Depth-First Search (DFS) algorithm, individual phase spaces are meticulously delineated. When given a specific phase transformation, the system can now pinpoint all corresponding compositions within the alloy system in real-time. The inherent versatility of this high-dimensional phase diagram, combined with AGI's acceleration potential, indicates a promising trajectory for its application across a myriad of various variables engineering tasks.

Such an encompassing reverse design methodology paves the way for a promising high-throughput alloy design paradigm. The synergy of the LCM's speed with the comprehensive nature of the high-dimensional phase diagram heralds a groundbreaking shift in materials science, propelling the field into a new era of innovation and discovery.

\section{Result and Discussion}
\label{sec:headings}

\subsection{Large CALPHAD model development}
The process of building the LCM is divided into two sub-tasks. As shown in Fig. 1A, the first step is dataset generation and the second step is the training of the machine learning model.

\subsubsection{Dataset generation}
In this work, the data was derived using the TC-toolbox, an API (Application Programming Interface) that interfaces MATLAB with Thermo-Calc. This combination permits automated, high-throughput calculations of phase volume of each phase across a diverse temperature and composition range. Temperature values were randomly sampled within the range of 673.15K to 2673.15K. Similarly, random compositions of Fe, Ni, Cr, and Mn were generated under the constraint that their sum totals to 1. This approach ensures a valid temperature and composition input. To guarantee adequate point density in the phase space, phase statuses of 1.2 million unique combinations of temperature and compositions are computed.

\subsubsection{Model training}
The model was designed to accept five inputs: T, x(Fe), x(Ni), x(Cr), and x(Mn). Given that there are total 15 kinds of phases in whole phase space, the model was set to produce outputs for 15 unique phase volumes. The index of outputs is corresponding to phase kinds. For instance, an index of '0' is associated with the volume of the ‘LIQUID’ phase, whereas an index of '14' pertains to the volume of the ‘SIGMA’ phase. Before the training process, temperature values underwent min-max normalization and any detected missing values were eliminated. The dataset was divided into a train set and a test set with a 4:1 ratio. A Backpropagation (BP) neural network with five hidden layers was utilized. The layers consist of 128, 256, 512, 256, and 15 neurons respectively. The ReLu activation function was employed for all layers. 

The uniqueness of the task, a hybrid task combining both multi-classification and regression —where the sum of phase volumes must equal 1—necessitated a tailored network to cater to this specific need. Rather than adjusting output data to align with standard neural network constraints, we incorporated an infrequently used last-layer Normalization activation function and employed Mean Square Error (MSE) as the loss function for this nuanced task. This approach ensures the output data conveys both volume and phase classification details. An indication of whether the predicted phase volume surpasses zero reveals the classification outcome.

The regular activation function in classification task, softmax function, defined as:
\begin{equation}
    softmax(x_{i})=\frac{e^{x_{i}}}{\sum_{j}^{}e^{x_{j}}} 
\end{equation}

ensures outputs sum to 1. However, due to the exponential nature of its formula, it tends to accentuate differences between input values. As illustrated in Fig. 1B, models employing softmax demonstrated a more gradual decline in validation loss in contrast to those using the normalize function, hinting at a potential compromised generalization capability with softmax. When the normalization function, defined as:
\begin{equation}
    Normalized(x_{i})=\frac{x_{i}}{\sum_{j}^{}x_{j}} 
\end{equation}
scales outputs to sum to unity, providing a balanced representation. For this study, it was favored over softmax for the last layer's activation, as evidenced by a more rapid decline in validation loss in Fig. 1B, suggesting superior generalization. 

\subsubsection{Model evaluation}
As shown in Fig. 1C and D, for the classification task, the accuracy achieved was 97.49\% on the training set and 97.40\% on the test set. In terms of phase volume prediction, the MSE was $4.75\times10^{-5}$ for the training set and $4.80\times10^{-5}$ for the test set, indicating an average volume prediction error of approximately 0.5\%. Such precision adequately meets the requirements for alloy design. To provide a visual representation of the model's accuracy, Fig. 1E presents the phase diagram generated by the machine learning model, while Fig. 1F is the phase diagram produced by Thermo-Calc console. The remarkable similarity between the two diagrams underscores the precision of the LCM in phase classification.

A substantial enhancement in processing speed was observed upon transitioning tasks from the CPU (running one Matlab client on Intel(R) Core(TM) i5-11300H) to the GPU (running LCM on NVIDIA RTX 3080 with 16GB video memory). As illustrated in Fig. 1G, the CPU achieved 0.8 points per second, whereas the GPU impressively delivered 39,800 points per second, highlighting the efficiency of GPU-accelerated computations.

\begin{figure}
  \centering
  \includegraphics[width=0.8\textwidth]{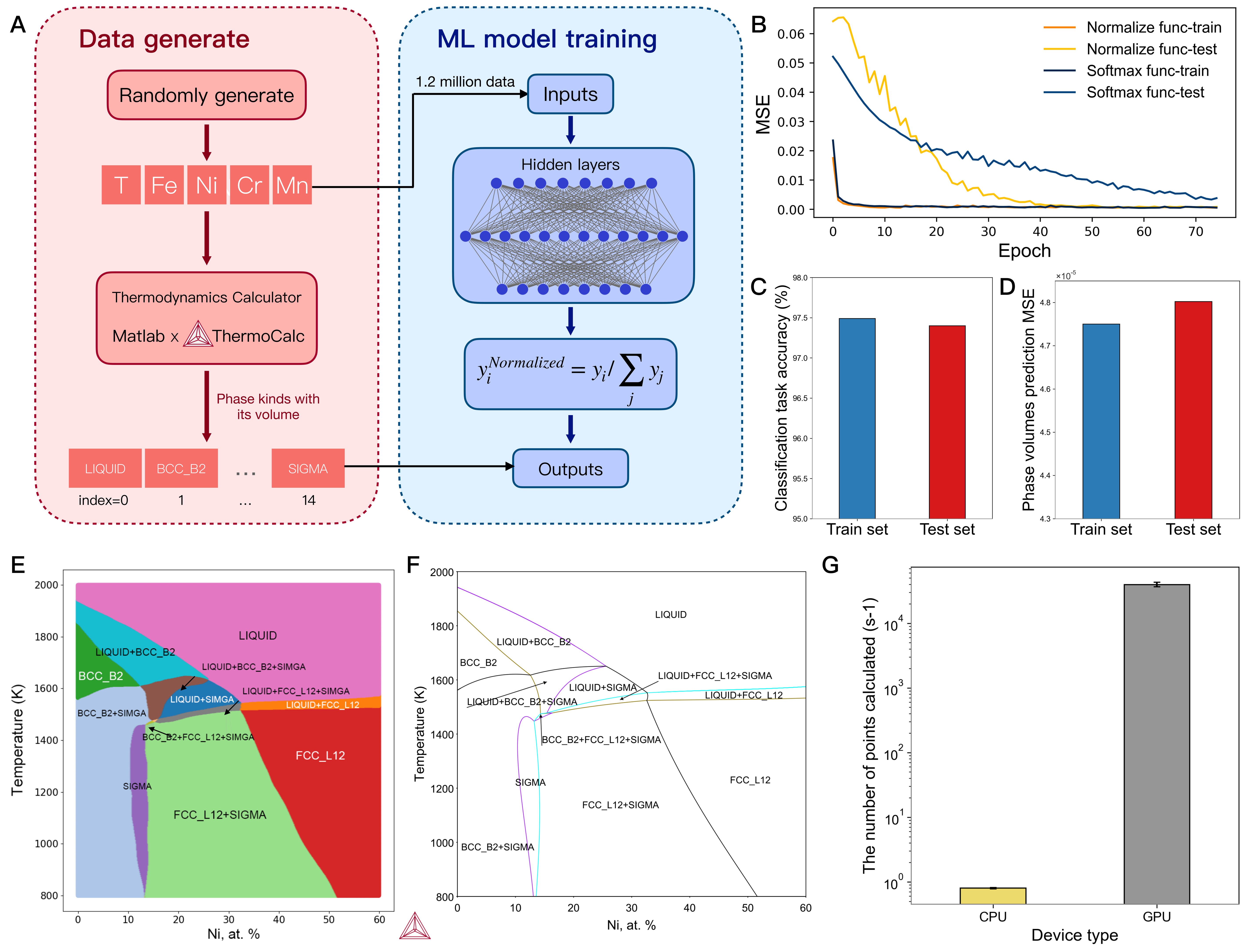}
  \caption{The establishment and validation of LCM. (A) The two sub-tasks involved in constructing LCM. (B) Comparison of the training process using the last layer activation functions: Softmax vs. Normalization. (C and D) Accuracy for the classification task (C) and MSE for phase volume prediction task (D). (E and F) Phase diagrams produced by LCM (E) and by the Thermo-Calc console (F) with x(Fe) and x(Mn) both fixed at 20\%. (G) Comparison of calculation speeds between CPU and GPU.}
  \label{fig:fig1}
\end{figure}

\subsection{High-dimensional phase diagram establishment}
The incorporation of LCM into alloy design has facilitated several advancements. Beyond merely accelerating existing alloy design algorithms, these models present opportunities for exploring new computational approaches. A prominent application arising from this integration is the development of high-dimensional phase diagrams. In essence, machine learning extends the scope and efficiency of traditional alloy design techniques.

In this study, to get the entire phase space data in FeNiCrMn system, the temperature range spanned from 673.15K to 2673.15K, with increments of 20K, giving 101 distinct values. With four elements in consideration, three independent variables arise. Factoring in temperature, four independent variables emerge.  For this research, T, x(Fe), x(Ni), and x(Cr) were chosen as these variables. Similarly, the compositions also had 101 possible values each with 1 at. \% increments. Yet, since x(Fe), x(Ni), and x(Cr) combined sum couldn't surpass 100\%, only 176,851 combinations are viable. Multiplied by the 101 temperature values, this yields a dense matrix of 17,861,951 valid data points characterizing the FeNiCrMn phase space. Significantly, the LCM calculated all these valid data points within 8min, while Matlab TC-toolbox may take several months to accomplish calculation.

Frequently, alloy design includes at least two conditions as the target phase transformation, for instance, a certain phase at higher temperatures and another at lower temperatures. Directly employing all 17,861,951 data points to pinpoint compositions that satisfy these conditions would involve a time complexity of O(n2), given that the search process incorporates two nested loops each traversing the dataset of n=17,861,951. Such a quadratic time complexity renders the task highly time-intensive. As depicted in Fig. 2A, the entire phase space was structured for efficient data access, harnessing a hashing mechanism.

With these discretized values, a multi-dimensional list, named phase\_space, was created with the dimensions 101x101x101x101. This arrangement, facilitated by hash mappings, enables direct access to data points, eliminating the need for loop-based searches. As an example, phase\_space[30][12][12][24] would specifically refer to the scenario where the temperature is 1273.15K, x(Fe) is 12\%, x(Ni) is 12\%, and x(Cr) is 24\%.

With the structured approach of the phase\_space list, while we can quickly access any data point, there still remains the challenge of identifying and categorizing continuous and distinct phase spaces within this high-dimensional dataset. In traditional alloy design using phase diagrams, researchers visualize phases as contiguous regions within a two-dimensional space. These diagrams predominantly feature composition on the x-axis and temperature on the y-axis. When studying these 2D diagrams, it becomes essential to identify overlapping areas across different temperatures to understand phase transformations. Such insights are crucial for alloy design, informing decisions about which compositions can exhibit desired phase characteristics over specific temperature ranges.

Transitioning from this two-dimensional representation to a more intricate four-dimensional or higher dimension space retains the fundamental principles of continuity and overlap. In this situation, continuous phase spaces need to be identified to provide insights into phase behaviors across different compositions and temperatures. However, mapping such continuous phase spaces in a high-dimensional space is complex, even in a ternary phase diagram with three dimensions(14). While two-dimensional continuity can be easily visualized on paper, the representation becomes abstract in four dimensions. To navigate and dissect this multi-dimensional space effectively, a systematic search algorithm is required. As shown in Fig. 2B, DFS serves this purpose, allowing for exploration of the four-dimensional space to identify the critical, continuous phase spaces. From a start point in this four-dimensional space, the target phase is determined. The algorithm seeks to navigate as far as possible along one dimension (or direction) before backtracking to explore alternate paths. To prevent revisiting points, they are marked as 'visited' after their initial encounter. This procedure mirrors maze-solving tactics, where one endeavors to traverse every pathway to its end before reverting to explore other untouched paths.

Considering the extensive nature of the dataset and the depth of potential searches, there's a risk of stack overflow with recursive methods. To mitigate this, a stack-based DFS approach was adopted.

In this four-dimensional list, eight distinct directions can be approached from any given point. By methodically iterating through these directions from a designated starting point, the DFS algorithm thoroughly explores the entire independent phase space linked to that origin.

To provide a clearer understanding, Fig. 2C visualizes this exploration. For simplification and visualization purposes, the temperature variable has been excluded, detailing the exploration process within the BCC\_B2 independent phase space through the DFS algorithm . It's also worth noting that, to minimize the effects of potential prediction errors, only continuous phase spaces encompassing more than 200 points were taken into account, the physical significance of the 200 points is the variation in each composition is about 3.7\% and temperature extent is 75K. The phase space less than these variations is considered meaningless for alloy design. In FeNiCrMn system, a total number of 51 different phase spaces which contain more than 200 points is found. Each continuous phase spaces data is saved independently so it can be easily used when design alloys with certain conditions.  Building upon this, the high-dimensional phase diagram is established.

\begin{figure}
  \centering
  \includegraphics[width=1\textwidth]{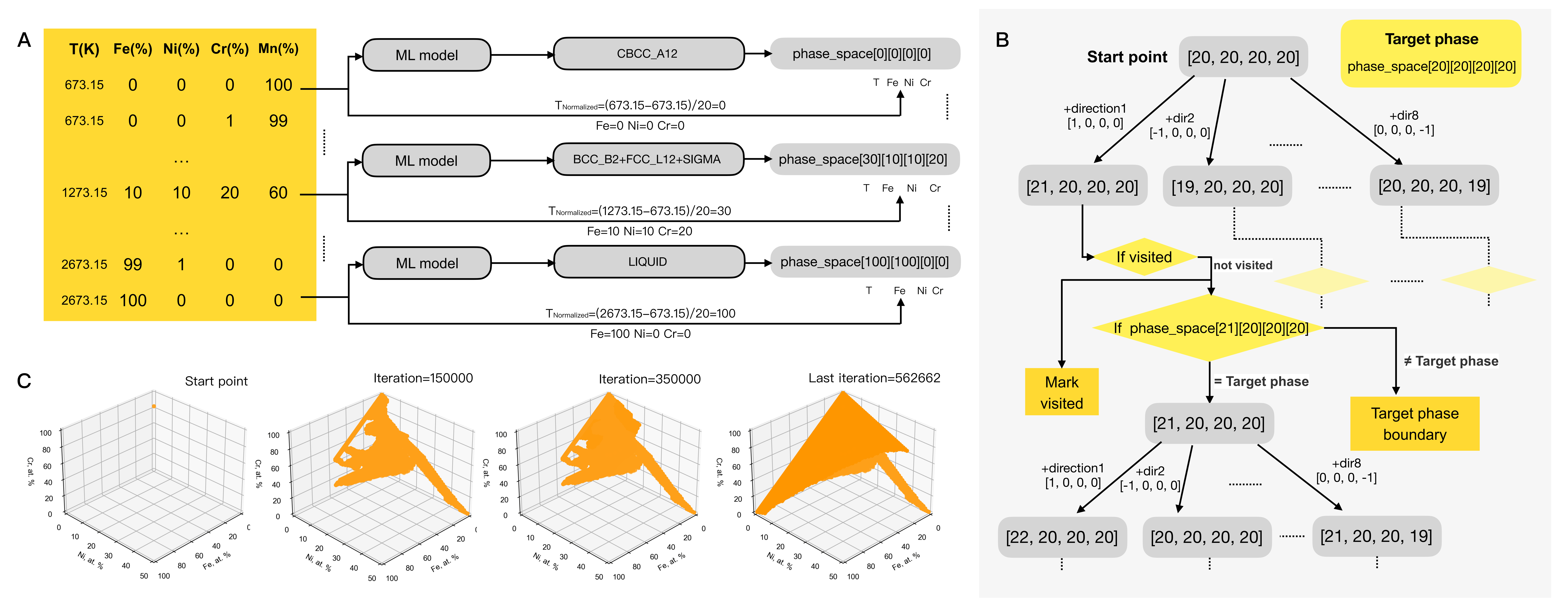}
  \caption{(A) Conversion of discretized data into a high-dimensional hash table termed 'phase\_space' list. (B) The process how DFS effectively discretizes each continuous phase space. (C) A visual representation of the DFS process identifying the BCC\_B2 phase space, with temperature omitted for clarity in visualization.}
  \label{fig:fig2}
\end{figure}

\subsection{‘Reverse design method’ leveraging High-dimensional phase diagrams}

The utilization of high-dimensional phase diagrams offers a distinct advantage in alloy design through the "reverse design method." Unlike traditional techniques where compositions are assessed sequentially in a point-by-point validation process, high-dimensional phase diagrams offer a more comprehensive approach. In the point-by-point validation method, each potential alloy composition is individually evaluated for its phase status. This can be a meticulous and prolonged procedure, frequently resulting in only a limited number of viable alloy compositions being identified. In contrast, high-dimensional phase diagrams adopt a reverse design approach: by specifying the desired targets first, they enable the comprehensive identification of all alloy compositions within a system that meet these prerequisites. This approach not only offers a holistic view but also streamlines the alloy design process, ensuring that no potential composition is overlooked.

To illustrate, when designing an alloy that possesses a single FCC\_L12 phase at higher temperatures and precipitates the MNNI\_L10 phase at cooler temperatures, one doesn't merely get a few alloy compositions. Instead, one can systematically explore overlaps between the FCC\_L12 phase space and the FCC\_L12 + MNNI\_L10 phase spaces to derive a comprehensive set of potential compositions.

Further, the integration of LCM accelerates the determination of phase volumes, enabling swift design iterations within particular temperature, composition, or phase volume constraints. Typically, each design task can be wrapped up in about 20s. Table. 1 catalogs all foundational alloy design varieties (eutectic, eutectoid, and precipitate) available within the FeNiCrMn system, emphasizing the exhaustive nature of this approach. 

\begin{table}[H]
 \caption{The number of foundational reactions within the FeNiCrMn system.}
  \centering
  \begin{tabular}{ccc}
    \toprule
    Reaction type     & Number of reactions   \\
    \midrule
    Eutectic reaction & 3     \\
    Eutectoid reaction  & 24    \\
    Precipitation reaction     & 22          \\
    \bottomrule
  \end{tabular}
  \label{tab:table}
\end{table}

\subsection{A showcase of the promising high-throughput alloy design paradigm.}
The recent unveiling of the high-dimensional phase diagram heralds a transformative shift in alloy design methodologies. No longer constrained to isolated compositions or a limited set of experiments, the new approach advocates for systematic, expansive experimentation. This allows for deeper and more comprehensive exploration of alloy systems. To illustrate this paradigm shift, the design of all eutectic alloys in the FeNiCrMn system is presented as a prime example. To ensure reliability, the temperature between the ‘LIQUID’ phase space and the eutectic phase space is under 20K, with the eutectic phase space spanning a vertical temperature range of at least 100K. Although this eutectic alloy design serves as a specific showcase, it embodies the expansive capabilities offered by the novel approach.

Fig. 4A illustrates the groundbreaking capability of this method. Every conceivable composition that qualifies as a eutectic alloy is systematically mapped. It's a testament to the vast landscape of possibilities the phase diagram reveals. In the past, reliance on heuristic approximations and iterative adjustments was common, but contemporary methodologies have advanced beyond these techniques. The future promises systematic exploration, with potential alloy compositions laid out in a comprehensive panorama.

To underscore the potential of the high-dimensional phase diagram in guiding high-throughput alloy design experiments, its outcomes are scrutinized against the established theoretical benchmarks of the CALPHAD method, as calculated by Thermo-Calc. The step figures from the Thermo-Calc TC-toolbox serve as a comparative yardstick, ensuring the diagram's reliability as a pivotal tool in the realm of alloy design. Fig. 4B-L presents 11 such step figures with linear variations in x(Fe), offering a systematic alloy design perspective. Eutectic reactions are highlighted using gradient colors, emphasizing the efficacy of the high-dimensional phase diagram as validated by the CALPHAD method in ensuring reliable, systematic alloy design. In FeNiCrMn system, 439 eutectic compositions are designed by high-dimensional phase diagram.

While this study emphasizes eutectic alloys, its ramifications span a broader spectrum. The high-dimensional phase diagram is not merely an analytical instrument but serves as a foundation for advanced high-throughput alloy design. It's well-suited for alloys with four or more elements and holds promise for diverse engineering tasks involving multiple variables. With this tool at our disposal, there's a gentle nudge towards a more systematic exploration of every designed alloy. When other properties are adapted to its AGI and high-dimensional phase diagrams, including DFT for physical property simulations(15), diffusion analyses(16), and FEM for mechanical simulations(9, 17), we could be on the brink of discovering materials with unprecedented and transformative characteristics.

\begin{figure}[H]
  \centering
  \includegraphics[width=1\textwidth]{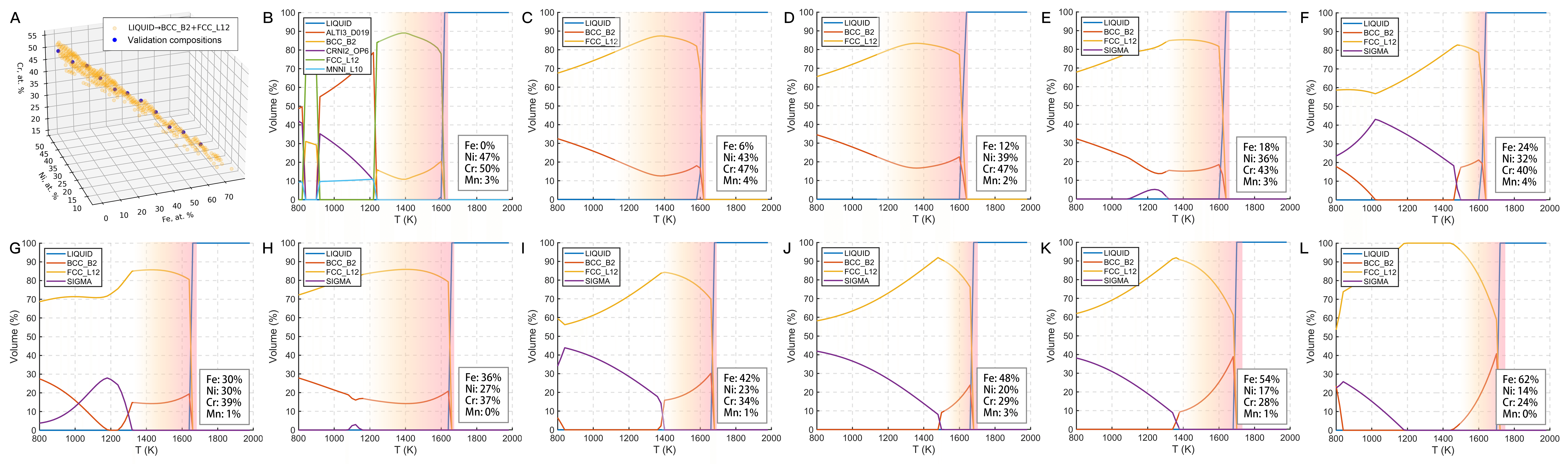}
  \caption{(A) All 439 eutectic reaction compositions in the FeNiCrMn system alongside their validation compositions. (B-L) A systematic alloy design perspective. 11 step figures drawn by TC-toolbox with linear x(Fe) variations.}
  \label{fig:fig3}
\end{figure}

\section*{Declaration of Competing Interest}
The authors declare that they have no known competing financial 
interests or personal relationships that could have appeared to influence 
the work reported in this paper.

\section*{Data availability}
Data will be made available on request.

\section*{Acknowledgments}
WWS gratefully acknowledges the support of National Natural Science Foundation for Young Scholars of China (Grant No. 52001063), National Natural Science Foundation of China (Grant No. 52371103), and Jiangsu Key Laboratory for Light Metal Alloys (Grant No. LMA202201). XLA gratefully acknowledges the support from Natural Science Foundation for Young Scholars of Jiangsu Province (Grant No. BK20220628), and National Natural Science Foundation for Young Scholars of China (Grant No. 52301130).

\section*{References}
\begin{enumerate}
    \item L. Kaufman, H. Bernstein, Computer calculation of phase diagrams with special reference to refractory metals. Academic Press Inc (1970).
    \item J. Du, V. Jindal, A. P. Sanders, K. S. Ravi Chandran, CALPHAD-guided alloy design and processing for improved strength and toughness in Titanium Boride (TiB) ceramic alloy containing a ductile phase. Acta Mater 171 (2019).
    \item A. Janz, J. Gröbner, H. Cao, J. Zhu, Y. A. Chang, R. Schmid-Fetzer, Thermodynamic modeling of the Mg-Al-Ca system. Acta Mater 57 (2009).
    \item S. Zhu, J. Shittu, A. Perron, C. Nataraj, J. Berry, J. T. McKeown, A. van de Walle, A. Samanta, Probing phase stability in CrMoNbV using cluster expansion method, CALPHAD calculations and experiments. Acta Mater 255 (2023).
    \item A. J. Zaddach, Zaddach, A. Joseph, Physical Properties of NiFeCrCo-based High-Entropy Alloys. PhDT (2015).
    \item W. W. Sun, R. K. W. Marceau, M. J. Styles, D. Barbier, C. R. Hutchinson, G phase precipitation and strengthening in ultra-high strength ferritic steels: Towards lean ‘maraging’ metallurgy. Acta Mater 130 (2017).
    \item E. Menou, G. Ramstein, E. Bertrand, F. Tancret, Multi-objective constrained design of nickel-base superalloys using data mining- and thermodynamics-driven genetic algorithms. Model Simul Mat Sci Eng 24 (2016).
    \item F. Tancret, Computational thermodynamics and genetic algorithms to design affordable $\gamma$'-strengthened nickeliron based superalloys. Model Simul Mat Sci Eng 20 (2012).
    \item X. Shang, Z. Liu, J. Zhang, T. Lyu, Y. Zou, Tailoring the mechanical properties of 3D microstructures: A deep learning and genetic algorithm inverse optimization framework. Materials Today, doi: https://doi.org/10.1016/j.mattod.2023.09.007 (2023).
    \item R. Kannan, P. Nandwana, Accelerated alloy discovery using synthetic data generation and data mining. Scr Mater 228 (2023).
    \item Y. Zeng, M. Man, K. Bai, Y. W. Zhang, Explore the full temperature-composition space of 20 quinary CCAs for FCC and BCC single-phases by an iterative machine learning + CALPHAD method. Acta Mater 231 (2022).
    \item P. Korotaev, A. Yanilkin, Steels classification by machine learning and Calphad methods. CALPHAD 82 (2023).
    \item C. Wang, W. Zhong, J. C. Zhao, Insights on phase formation from thermodynamic calculations and machine learning of 2436 experimentally measured high entropy alloys. J Alloys Compd 915 (2022).
    \item A. Abu-Odeh, E. Galvan, T. Kirk, H. Mao, Q. Chen, P. Mason, R. Malak, R. Arróyave, Efficient exploration of the High Entropy Alloy composition-phase space. Acta Mater 152 (2018).
    \item Z. Rao, P. Y. Tung, R. Xie, Y. Wei, H. Zhang, A. Ferrari, T. P. C. Klaver, F. Körmann, P. T. Sukumar, A. K. da Silva, Y. Chen, Z. Li, D. Ponge, J. Neugebauer, O. Gutfleisch, S. Bauer, D. Raabe, Machine learning–enabled high-entropy alloy discovery. Science (1979) 378 (2022).
    \item J. Zhong, W. Chen, L. Zhang, HitDIC: A free-accessible code for high-throughput determination of interdiffusion coefficients in single solution phase. CALPHAD 60 (2018).
    \item R. Quey, M. Kasemer, The Neper/FEPX Project: Free / Open-source Polycrystal Generation, Deformation Simulation, and Post-processing. IOP Conf Ser Mater Sci Eng 1249 (2022).
\end{enumerate}

\end{document}